\documentstyle[prd,aps,twocolumn]{revtex}
\begin{document}
\draft

\def\slash#1{\rlap{\hbox{$\mskip 1 mu /$}}#1}   
\def\bra#1{\left\langle #1\right|}              
\def\ket#1{\left| #1\right\rangle}              

\twocolumn[\hsize\textwidth\columnwidth\hsize\csname
@twocolumnfalse\endcsname

\begin{flushright}
ANL-HEP-PR-00-093\\
\end{flushright}

\title{Compatibility of various approaches to heavy-quark fragmentation}
\author{G.~T.~Bodwin and B.~W.~Harris}
\address{High Energy Physics Division,
             Argonne National Laboratory,
             Argonne, Illinois 60439 }
\date{December 2000}

\maketitle

\vspace*{0.2in}
\begin{abstract}
We find that the definition of the heavy-quark fragmentation function
given by Jaffe and Randall differs by a factor of the
longitudinal-momentum fraction $z$ from the standard Collins-Soper
definition. Once this factor is taken into account, the explicit
calculation of Braaten {\em et al.}\ is found to be in agreement with
the general analysis of Jaffe and Randall. We also examine the model of
Peterson {\em et al.}\ for heavy-quark fragmentation and find that the
quoted values of the width and of the value of $z$ at the maximum are in
error. The corrected values are in agreement with the analysis of Jaffe
and Randall.
\end{abstract}

\vspace*{0.2in}
\vskip1.0pc]

\section{Introduction}

There does not yet exist a complete first-principles calculation of the
nonperturbative transition of a parton (emerging from a high energy
scattering process) to a hadron. Nevertheless, a rigorous formal
description of fragmentation functions in terms of operator matrix
elements has been available for some time \cite{collins}. Additionally,
fragmentation functions are an integral part of present day
phenomenology.

When the parton involved in the fragmentation is a relatively heavy
charm or beauty quark, there are simplifications in the treatment of
fragmentation functions that allow one to make additional theoretical
progress. It is the case of heavy-quark fragmentation that we consider
here.

A theoretical description of fragmentation that is based on the
operator-matrix-element definition of the fragmentation function in 
heavy-quark effective theory (HQET) \cite{hqet} has been presented by
Jaffe and Randall \cite{jaffe}.  They show that the heavy-quark
expansion of the fragmentation function must take a specific form that
depends only on certain combinations of variables. In this paper, we
point out that the Jaffe-Randall definition of the fragmentation
function differs from the standard Collins-Soper definition by a factor
of the longitudinal-momentum fraction $z$ and that this difference is
crucial to the proper interpretation of the Randall-Jaffe form.

The work of Braaten {\em et al.}\  \cite{braaten} models the heavy-quark
fragmentation function with a fixed-order perturbative calculation in
the context of HQET. The results of this calculation appear, at first
sight, to be incompatible with the Jaffe-Randall form. However, after
accounting for the difference between the Jaffe-Randall definition of
the fragmentation function and the Collins-Soper definition, which is
used by Braaten {\em et al.}, we find agreement between the explicit
calculation and the Jaffe-Randall form.

Perhaps the most widely-used parameterization for fragmentation
functions is that of the model of Peterson {\em et al.}\ \cite{peterson},
which is based on old-fashioned perturbation theory. The Peterson
parameterization has been fit to charm and beauty production data
\cite{fit} and subsequently used as input in making predictions for
hadron-, photo-, and electro-production of charm and beauty \cite{pre}.

It has been suggested that the Peterson {\em et al.}\ parameterization
is incompatible the Jaffe-Randall form.  In particular, it has been
reported that the width of the fragmentation function is different in
the two approaches \cite{jaffe}. We find that the width and value of $z$
corresponding to the maximum are reported incorrectly in the original
Peterson {\em et al.}\ paper \cite{peterson}. The corrected values are
compatible with the Jaffe-Randall form.

In Sec.~II, we give an overview of the general HQET analysis of Jaffe
and Randall. We discuss the compatibility of the explicit calculation of
Braaten {\em et al.}\ with the Jaffe-Randall form in Sec.~III. Sec.~IV
contains a description of the model of Peterson {\em et al.}\ and a
discussion of its relation to the Randall-Jaffe form.  Our findings are
summarized in Sec.~V.

\section{The HQET Analysis of Jaffe and Randall}

We now discuss the work of Jaffe and Randall \cite{jaffe}, which
provides a QCD-based interpretation of heavy-quark fragmentation in
terms of the heavy-quark mass expansion. We begin with the standard
Collins-Soper \cite{collins} definition of the fragmentation function
for a heavy quark into a heavy hadron:
\vfill\eject
\begin{eqnarray}
\hat f(z,\mu^2) &=& \frac{z^{d-3}}{4N_c} \biggl[\int {d\lambda\over 2\pi}
e^{i \lambda/z} {\rm Tr}\, \slash{n} \nonumber\\
& \times &
\bra{0} h(\lambda n)\ket{H'(P)}\bra{H'(P)} \overline h(0) \ket{0}\biggr],
\label{op}
\end{eqnarray}
where the trace is over color and Dirac indices, $N_c$ is the number of
colors, $h(x)$ is the heavy-quark field at space-time position $x$, $P$
is the four-momentum of the heavy hadron, $n$ is defined by $n^2=0$ and
$n\cdot P=1$, and $z=n\cdot P/n\cdot k$, where $k$ is the heavy-quark
momentum. The state $\ket{H'(P)}$ consists of the heavy hadron plus any
number of additional hadrons. The matrix element is understood to be
evaluated in the light-cone gauge $n\cdot A=0$. Eqn.~(\ref{op}) is valid
in $d$ dimensions. However, in discussing the Jaffe-Randall analysis, we
specialize to the case of four dimensions.

The definition (\ref{op}) is normalized such that $\hat f(z,\mu^2)$
times the parton-level cross section reproduces the full QCD cross
section, differential in $z$, in the collinear limit. By ``collinear
limit'' we mean the limit in which $\sqrt{k^2}$ and the transverse
component (relative to $P$) of the 3-momentum of $k$ are neglected in
comparison to the longitudinal component of the 3-momentum of
$k$.\footnote{
The normalization on the right side of Eq.~(\ref{op}) can be understood
as follows. For simplicity, we consider a frame in which
$P=(P^+,P^2/P^+,0_\bot)$ has no transverse component. $k=(k^+,k^-,k_\bot)$,
with $P^+=zk^+$. In this frame, $n=(0,1/P^+,0_\bot)$ and
$\lambda=P^+x^-$, where $x$ is the space-time separation of the quark
and antiquark fields. In the collinear limit, $k^-$ and $k_\bot$ are
neglected in comparison with $k^+$. In that limit, the full QCD cross
section is reproduced by the parton-level cross section, times the
Fourier transform in brackets in Eq.~(\ref{op}), times a normalization
factor. This normalization factor includes $1/(2P^+)$ for the
normalization of the hadron state, $2k^+$ for the normalization of the
quark state in the parton-level cross section, $1/4$ from the Fierz
re-arrangement $I_{\alpha\alpha'}I_{\beta\beta'}\rightarrow
(1/4)\gamma^+_{\alpha\beta}\gamma^-_{\alpha'\beta'}$ that factors the
Dirac indices, $1/N_c$ from the Fierz re-arrangement
$I_{aa'}I_{bb'}\rightarrow (1/N_c)I_{ab}I_{a'b'}$ that factors the color
indices, and $1/k^+$ for the quark sum-over-spinors projector
($\gamma\cdot k\approx\gamma^- k^+$) in the parton-level cross section.
The full QCD cross section is differential in the longitudinal momentum
of the final-state hadron. Since $dP^+=-k^+dz$, we obtain an
additional factor of $k+$ on conversion to a cross section that is
differential in $z$. In the full QCD cross section, there is also an
integral over the transverse momentum of the hadron. This integration
corresponds, in the frame in which the hadron has zero transverse
momentum, to an integration with respect to $d^{d-2}(zk_\bot)$ 
(Ref.~\cite{collins}). The conversion to an integration with respect to
$d^{d-2}k_\bot$ yields a factor $z^{d-2}$. (The integration over all
$k_\bot$ corresponds to taking the quark and antiquark fields at
transverse separation $x_\bot=0$.) Altogether, we obtain a factor
$z^{d-3}/(4N_c)$, which is just the coefficient of the Fourier transform
in Eq.~(\ref{op}).} This normalization ensures that $\hat f(z,\mu^2)$
is a probability distribution in $z$ (Ref.~\cite{collins}).

We note that the definition (\ref{op}) contains a factor $z$ (in four
dimensions) relative to the definition of the fragmentation function
used by Jaffe and Randall \cite{jaffe}. This factor $z$ will be
important in comparing with the work of Braaten {\em et al.}\ below.

The analysis of Eq.~(\ref{op}) in HQET proceeds as follows. First,
following the standard method for obtaining the heavy-quark-mass
expansion, one decomposes the field $h(x)$ into the sum of a large 
component $h_v(x)$ and a small component $\underline{h}_v(x)$:
\begin{eqnarray}
h_v(x) &=& e^{-im_Q v \cdot x} P_+ h(x) \\
\underline{h}_v(x) &=& e^{-im_Q v \cdot x} P_- h(x),
\end{eqnarray}
with $m_Q$ the heavy-quark mass, $P_\pm = (1\pm \slash{v})/2$, and $v$
the hadron's four-velocity. The leading term in the mass expansion of
$f(x,\mu^2)$ is contained in the large-large combination of fields:
\begin{eqnarray}
\hat f(z,\mu^2)&=&\frac{z}{4N_c} \int {d\lambda \over 2\pi}e^{i \lambda/z}
{\rm Tr}\, \slash{n} e^{-im_Q\lambda n \cdot v} \nonumber\\
& \times & \bra{0} P_+h(\lambda n)\ket{H'(P)}\bra{H'(P)}
\overline{P_+h(0)} \ket{0} \nonumber \\
&+&\ldots\; .
\label{expan}
\end{eqnarray}
Here, and throughout this paper, we use the ellipsis to denote terms of 
higher order in the hadron-mass expansion.

In Ref.\ \cite{jaffe}, it is argued that the matrix element in
Eq.~(\ref{expan}) is a dimensionless function ${\cal F}(\lambda
\delta)$. This function may be written in terms of its Fourier
transform:
\begin{equation}
{\cal F}(\lambda \delta) = 2 \int_{-\infty}^{\infty} d\alpha
e^{-i\alpha \lambda \delta} \, a(\alpha) \, ,
\label{fourier}
\end{equation}
where 
\begin{equation}
\delta=1-m_Q/m_H,
\label{delta}
\end{equation}
and $m_H$ is the hadron mass. Inserting Eq.\ (\ref{fourier}) into Eq.\
(\ref{expan}), one can evaluate the integral, with the result
\begin{equation}
\hat f(z,\mu^2)=\frac{z}{\delta} \hat a \left( \frac{1/z-m_Q/m_H}
{\delta} \right)+\ldots \; .
\end{equation}
A more complete analysis in Ref.~\cite{jaffe} also yields the
next-to-leading term in the hadron-mass expansion:
\begin{equation}
\hat{f}(z,\mu^2) = z\left[\frac{1}{\delta} \hat{a}(y)
+ \hat{b}(y) + \cdots \right] \, ,
\label{rj1}
\end{equation}
where $y=(1/z-m_Q/m_H) / \delta$. The analysis in Ref.~\cite{jaffe}
does not yield a precise prediction for the functional form of $a$ and
$b$, but some general properties may be deduced. The function $a$
describes, in the limit of infinite heavy-quark mass, the effects of
binding in the heavy hadron on the heavy-quark momentum distribution. For
a free heavy quark, $a(y)$ would be a $\delta$-function at $y=1$. In a
heavy hadron, the binding smears the heavy-quark momentum distribution.
It can be shown \cite{jaffe} that the distribution has a width
\begin{mathletters}
\begin{equation}
\Delta z\sim \delta
\end{equation}
and a maximum at $z=z_{\rm max}$, where
\begin{equation}
1-z_{\rm max} \sim \delta.
\end{equation}
\label{j-r-shape}
\end{mathletters}%

\section{A perturbative model}

Braaten {\em et al.}\ \cite{braaten} present a QCD-inspired model for
the fragmentation of a heavy quark into an $S-$wave light-heavy meson.
In this model, the fragmentation function is computed in perturbative
QCD (at Born level) in an expansion in inverse powers of the heavy-quark
mass.

Braaten {\em et al.}\ define the fragmentation function as the collinear
limit of the ratio of the cross section for producing a hadron to the
cross section for producing a quark.  As we explained in Sec.~II, this
definition is equivalent to the Collins-Soper definition (\ref{op}) and
leads to a fragmentation function that is a probability distribution in
$z$.

For the projection of the $Qq$ state onto the meson, Braaten {\em et al.}\ 
take the standard nonrelativistic-bound-state expression.  For
example, in the case of a $^1S_0$ meson, they assume the Feynman rule
for the $QqH$ vertex to be
\begin{equation}
\frac{\delta_{ij}}{\sqrt{3}} \frac{R(0)\sqrt{m_H}}{\sqrt{4\pi}}
\gamma_5 (1+\slash{v})/2,
\end{equation}
where $R(0)$ is the radial wave function at the origin. From the
terms of leading order in the heavy-quark mass expansion, they obtain,
in the case of a $^1S_0$ meson,
\begin{eqnarray}
\label{b1}
\hat f & \approx & N\left[{1\over \delta}\frac{(1-y)^2}{y^6} (3y^2+4y+8)
\right. \nonumber \\
& - & \left. \frac{(1-y)^3}{y^6} (3y^2+4y+8)\right],
\end{eqnarray}
where $N=2\alpha_s^2|R(0)|^2/(81\pi m_q^3)$.

At first glance, this result may seem to contradict the Jaffe-Randall
analysis, which shows that the terms of leading order in the heavy-quark
mass expansion give a contribution that is contained entirely in the
function $a(y)$ in Eq.~(\ref{rj1}). However, the factor $z$ in the
definition of the fragmentation function (\ref{op}) is crucial here.
From the definitions of $y$ and $\delta$, we have $z=1/[1-\delta(1-y)]$,
and, so, we can re-write Eq.~(\ref{b1}) as
\begin{equation}
\hat f/z\approx {N\over \delta}\frac{(1-y)^2}{y^6} (3y^2+4y+8),
\end{equation}
which is of the form of $a(y)$ in Eq.~(\ref{rj1}).

\section{Peterson Fragmentation}

Finally, we examine the model of Peterson {\em et al.}\ \cite{peterson}
for the fragmentation of a fast-moving heavy quark $Q$ with mass $m_Q$ into
a heavy hadron $H$ (consisting of $Q\overline{q}$) with mass $m_H$ and a
light quark $q$ with mass $m_q$. The basic assumption in this model is
that the amplitude for the fragmentation is proportional to $1/(\Delta
E)$, where $\Delta E = E_H + E_q - E_Q$ is the energy denominator for
the process in old-fashioned perturbation theory. It follows that the
probability for the transition $Q \rightarrow H + q$ is proportional to
$1/(\Delta E)^2$. Taking the heavy quark's momentum to define the
longitudinal axis, one can express $\Delta E$ in terms of the
magnitude of the heavy quark's momentum $P_Q$, the fraction $z$ of the
heavy quark's momentum that is carried by the heavy hadron, and the
transverse momentum $p_\bot$ of the heavy hadron or the light quark:
\begin{eqnarray}
\Delta E &=& \sqrt{m_H^2+p_\bot^2+z^2P_Q^2} + \sqrt{m_q^2+p_\bot^2
+(1-z)^2P_Q^2} \nonumber \\ &-& \sqrt{m_Q^2+P_Q^2} \nonumber \\
&\approx& \frac{m_H^2+p_\bot^2}{2zP_Q} +
\frac{m_q^2+p_\bot^2}{2(1-z)P_Q}
- \frac{m_Q^2}{2P_Q} + \cdots \nonumber \\
&\approx & -{m_Q^2\over 2P_Q}[1 - 1/z - \epsilon/(1-z)].
\label{energyd}
\end{eqnarray}
In the last line, we have set $m_H\approx m_Q$ and neglected $p_\bot^2$
relative to $m_Q^2$ and used the definition 
\begin{equation}
\epsilon \equiv (m_q^2+p_\bot^2)/m_Q^2.
\label{epsilon}
\end{equation}
Multiplying $1/(\Delta E)^2$ by a factor $1/z$ for the longitudinal
phase space, one arrives at the following {\it ansatz} for the
fragmentation function \cite{peterson}
\begin{equation}
D_Q^H(z) = \frac{N}{z[1 - 1/z - \epsilon/(1-z)]^2},
\end{equation}
where the normalization $N$ is fixed by the condition
\begin{equation}
\sum_H \int dz D_Q^H(z) = 1,
\end{equation}
and the sum extends over all hadrons that contain $Q$. 

Contrary to the claims in Ref.\ \cite{peterson}, we find that $D_Q^H(z)$
has a maximum at $z=z_{\rm max}$, with $z_{\rm max} \approx 1 -
\sqrt\epsilon$, and a width of order $\sqrt\epsilon$.  Specifically, we
find, that the distance between the inflection points of $D_Q^H(z)$ is
$[(8-2\sqrt{3})/3]^{1/2}\sqrt{\epsilon}$, and the full width at half
maximum is $2 \sqrt{\epsilon}$. In these values of $z_{\rm max}$ and the
width, we have neglected terms of higher order in $\epsilon$.

We can compare the Peterson {\em et al.}\ form with the Randall-Jaffe 
results by making use of Eqs.~(\ref{delta}) and (\ref{epsilon}), to 
obtain 
\begin{equation}
\sqrt{\epsilon}\approx \delta.
\label{parameters}
\end{equation}
It has been believed that the shape of the Peterson {\em et al.}\
fragmentation function is incompatible with results obtained from heavy
quark effective theory (HQET). However, it follows immediately from
Eq.~(\ref{parameters}) that our results for the width and $z_{\rm max}$ of
the Peterson form are compatible with the Jaffe-Randall constraints
(\ref{j-r-shape}).

\section{Summary}

We have examined the Randall-Jaffe analysis of the heavy-quark
fragmentation function in HQET and found that the Randall-Jaffe
definition of the heavy-quark fragmentation function differs from the
standard Collins-Soper definition (\ref{op}) by a factor $z$. This
factor is crucial to the interpretation of the fragmentation function as
a probability distribution in $z$. We have found that, once this factor
has been taken into account, the explicit, perturbative model
calculation of Braaten {\em et al.}\ is in agreement with the
Randall-Jaffe analysis.

We have also examined the model of Peterson {\em et al.}\ for the
heavy-quark fragmentation function. Our results for the width and the
value of $z$ corresponding to the maximum of the Peterson fragmentation
function differ from the values stated in the Peterson {\em et al.}\
paper \cite{peterson}. Our values are consistent with constraints from
the general analysis of Randall and Jaffe.

\acknowledgements
We thank R.~Jaffe for confirming the relationship of the Randall-Jaffe
definition of the fragmentation function to the Collins-Soper
definition. This work was supported by the United States Department of
Energy, High Energy Physics Division, under contract W-31-109-Eng-38.

\end{document}